\documentclass[twocolumn,preprintnumbers,pra,superscriptaddress,showkeys]{revtex4}

\usepackage{amsmath,amssymb,amsfonts,bbm,graphicx,float,times,psfrag}
\usepackage[pdftex]{color}
\usepackage{amsmath,bm}
\usepackage[colorlinks, linkcolor=blue, citecolor=blue,  breaklinks=true]{hyperref}
\usepackage{mathtools,amsfonts,mathptmx}
\usepackage[utf8]{inputenc}
\usepackage[T1]{fontenc}
\usepackage{xcolor}
\usepackage{braket}
\usepackage[titletoc,title]{appendix}
\usepackage[nameinlink,capitalize]{cleveref}

\begin{document}
\title{Synthetic magnetism enhanced mechanical squeezing in Brillouin optomechanical system} 

\author{D. R. Kenigoule Massembele}
\affiliation{Department of Physics, Faculty of Science, 
University of Ngaoundere, P.O. Box 454, Ngaoundere, Cameroon}

\author{P. Djorwé}
\email{djorwepp@gmail.com}
\affiliation{Department of Physics, Faculty of Science, 
University of Ngaoundere, P.O. Box 454, Ngaoundere, Cameroon}
\affiliation{Stellenbosch Institute for Advanced Study (STIAS), Wallenberg Research Centre at Stellenbosch University, Stellenbosch 7600, South Africa}

\author{Souvik Agasti}
\affiliation{IMOMEC division, IMEC, Wetenschapspark 1, B-3590 Diepenbeek, Belgium}
\affiliation{Institute for Materials Research (IMO), Hasselt University,	Wetenschapspark 1, B-3590 Diepenbeek, Belgium}

\author{K.S. Nisar}
\affiliation{Department of Mathematics, College of Science and Humanities in Alkharj, Prince Sattam Bin Abdulaziz University, Alkharj 11942, Saudi Arabia}

\author{A.K. Sarma}
\email{aksarma@iitg.ac.in}
\affiliation{Department of Physics, Indian Institute of Technology Guwahati, Guwahati 781039, India}

\author{A.H. Abdel-Aty}
\affiliation{Department of Physics, College of Sciences, University of Bisha, Bisha 61922, Saudi Arabia}

\begin{abstract}
We propose a scheme to generate large amount of mechanical squeezing, far beyond the $\rm{3dB}$ limit, which is based on synthetic magnetism in optomechanical system that hosts a Backward Stimulated Brillouin Scattering (BSBS) process. Our benchmark system consists of an acoustic mode coupled to two optical modes through the BSBS process, and a Duffing mechanical oscillator that couples to the same optical modes through the standard optomechanical radiation pressure. The synthetic magnetism comes from the modulation of the mechanical coupling between the acoustic and the mechanical mode. When there is no synthetic magnetism, a given amount of mechanical squeezing is generated in the system. This squeezing is mainly dependent on the BSBS process, and it is fragile against thermal noise. By switching on the synthetic magnetism, the degree of the generated squeezing is greatly enhanced and goes far beyond the limit of the $\rm{3dB}$. This large magnetism induced squeezing persists even when there is no BSBS process in the system. Moreover, this generated squeezing is robust enough against thermal noise in comparison to the one induced when the  synthetic magnetism is off. Furthermore, both the mechanical variance  squeezing and effective phonon number exhibit series of peaks and dips depending on the phase modulation of the mechanical coupling. This oscillatory feature is reminiscent of a sudden death and revival of squeezing phenomenon, which can be used to maintain a desired magnitude of squeezing by tuning this phase. Our proposal provides a path toward a flexible scheme that generates large amount of squeezing, far beyond the $\rm{3dB}$ limit.  Such a generated squeezed states can be used for quantum applications including quantum information processing, quantum sensing and metrology, and quantum computing.  
\end{abstract}

\maketitle
\date{\today}

\section{Introduction}\label{intro}
Nonclassical states such as squeezed and entangled states are crucial ingredients required to improve a range of quantum applications including quantum information processing \cite{Wendin_2017,Slussarenko_2019,Meher_2022}, quantum sensing and metrology \cite{Degen_2017,Polino_2020,Stray_2022,Barbieri_2022}, quantum computing \cite{Zidan_2021}, and quantum supremacy \cite{Arute_2019,Zhong_2020,Zhong_2021}. For instance, entangled states have been widely generated in optomechanical systems \cite{Aspelmeyer.2014,Foulla_2017} by exploring diverse techniques \cite{Riedinger_2018,Tchodimou.2017,Kotler_2021,Chakrabo.2019,Rostand1.2024}, and they were recently proposed as resources to enhance sensing \cite{Xia_2023,Brady_2023}. Beside that, large intracavity squeezed field have been generated \cite{Vahlbruch_2016,Qin_2022}, which can be used for mechanical cooling \cite{Clark_2017,Djor.2012}, gravitational wave detection \cite{Aasi_2013,Abbott.2016}, and to further generate nonclassical states \cite{Qin_2021}. Similarly, strong mechanical squeezed states \cite{Banerjee_2023} were equally generated, which can be useful to enhance mass-sensing \cite{Djorwe.2019,Tchounda_2023,djorwe.PRR}, 
information processing \cite{Stannigel_2012,Madiot_2023}, and state transfer \cite{Ren_2022} involving phonons. Despite these interesting applications related to squeezed states, their generation is often limited by quantum noise that resists any measurement below the Zero Point Fluctuation (\rm{ZPF}), i.e., beyond the $\rm{3dB}$ limit. The existing systems in which mechanical squeezing beyond $\rm{3dB}$ has been achieved have explored reservoir engineering technique or two-tone driving \cite{Woolley_2014,Wollman_2015,Lei_2016}.    

Recently, technique based on synthetic magnetism (engineered via a modulation of the photon/phonon hopping rate) have been used in optomechanical structures for specific purposes. For instance, an artificial magnetic field for photons was engineered to achieve photon transport as reported in \cite{Schmidt.2015,Fang_2012}. 
Similarly, synthetic magnetic field for phonon/acoustic has been created  \cite{Brendel.2017, Mathew_2020,Wang_2020} for  phononic transport at the nanoscale. More recently, synthetic magnetism has been engineered in optomechanical systems to enhance entangled states generation \cite{Lai_2022}. Owing to these interesting physics fostered by synthetic magnetism in optomechanics, here we use it to enhance mechanical squeezing in optomechanical system involving Backward Stimulated Brillouin Scattering (BSBS) process which has been proposed in \cite{BSBS_basis}.   

The underlying system, we are considering, consists of an acoustic (mechanical) mode coupled to two optical modes through the BSBS process (radiation pressure coupling). We propose a scheme to connect the acoustic and the mechanic modes through a mechanical coupling having a strength $J_m$ that is modulated via a phase $\theta$. Such a phase  modulation of the coupling induces a synthetic magnetism in our model system \cite{Lai_2022,Djor.2023}.  When the synthetic magnetism is switched off, a given amount of mechanical squeezing is generated in the system. This squeezing is mainly dependent on the BSBS process, and it is fragile against thermal noise \cite{BSBS_basis}. By switching on the synthetic magnetism, we found that i) the degree of the generated squeezing is greatly enhanced and goes far beyond the limit of the $\rm{3dB}$, ii) this squeezing persists even when there is no BSBS process in the system, and iii) this generated squeezing is robust enough against thermal noise in comparison to the case when the  synthetic magnetism is off. Furthermore, both the mechanical squeezing and effective phonon number exhibit series of peaks and dips depending on the phase modulation of the mechanical coupling. This oscillatory feature induces a sudden death and revival of squeezing, which can be used to maintain a desired magnitude of squeezing by tuning the phase. Our proposal paves a way towards a flexible scheme that can be used to generate arbitrary amount of squeezing. Such a generated squeezed states can be used for quantum applications including quantum information processing, quantum sensing/metrology, and quantum computing.  

The rest of our work is organized as follow. \autoref{sec:model} provides the dynamical equations and derives the analytical expressions involved in our investigation. The squeezing enhancement, together with the important results are presented throughout \autoref{sec:squeez}. Our work in concluded in \autoref{sec:concl}.

\section{Model and dynamical equations} \label{sec:model}

We consider a system consisting of an acoustic mode ($b_a$) that couples to two optical modes through the BSBS process, and a nonlinear mechanical oscillator ($b_m$) that couples also to the two optical modes. In order to enhance the squeezing of the mechanical oscillator, the acoustic and the mechanical modes are mechanically coupled, having coupling strength $J_m$. The coupling is modulated through a phase $\theta$ that induces a synthetic gauge into the dynamics. The Hamiltonian of such a system is given by ($\hbar=1$):
\begin{equation}\label{eq:eq1}
 H =H_{\rm{0}}+H_{\rm{OM}}+H_{\rm{BSBS}}+H_{\rm{int}}+H_{\rm{drive}}, 
 \end{equation}
where
\begin{eqnarray}  \label{eq:eq2}
H_{\rm{0}}&:=&\sum_{j=1,2} \omega_{c_j} a_j^\dagger  a_j + \omega_a b_a^\dagger b_a + \omega_m b_m^\dagger b_m, \\
H_{\rm{OM}}&:=&\sum_{j=1,2}g_{a,m} a_j^\dagger a_j   (b_m +b_m^\dagger)+  \frac{\eta}{2}(b_m + b_m^\dagger)^4, \\
H_{\rm{BSBS}}&:=&- g_a( a_1^\dagger a_2 b_a+ a_1 a_2^\dagger b_a^\dagger),\\
H_{\rm{int}}&:=&{J_m}({e^{i\theta}}{b_{a}^{\dagger}}{b_m} + {e^{-i\theta}}{b_a}{b_{m}^{\dagger}})\\
H_{\rm{drive}}&:=&\sum_{j=1,2}iE_j(a_j^{\dagger }e^{-i\omega_{p_j} t} - a_je^{i\omega_{p_j} t}).
\end{eqnarray} 
In the above Hamiltonian, the first term $H_{\rm{0}}$ is the free Hamiltonian corresponding to the optical ($a_j$), the acoustic ($b_a$)  and the mechanical ($b_m$) mode. The first term in $H_{\rm{OM}}$ captures the optomechanical interaction between the optical and the mechanical modes, while the second one accounts for nonlinear effect on the mechanical resonator through the Duffing coefficient $\eta$.  The third term $H_{\rm{BSBS}}$ stands for the triply resonant phonon-photon interaction triggered via the $\rm{BSBS}$  process. $H_{\rm{int}}$ describes the mechanical coupling between the acoustic and the mechanical modes. The single-photon optomechanical (Brillouin) coupling, $g_m$ ($g_a$), results from the radiation pressure (electrostrictive) force. The last term,  $H_{\rm{drive}}$, denotes the drivings fields. Here $E_j$ and $\omega_{p_j}$ are the amplitude and the frequency of the $j^{th}$ field. In what follows, we will assume the same frequency field, $\omega_{p_{1,2}}\equiv\omega_{p}$. The optical cavity frequency are $\omega_{c_j}$ and the mechanical (acoustic) frequency is $\omega_m$ ($\omega_a$).
In the frame rotating at $H_r=\omega_{p_1}a_1^\dagger a_1 +\omega_{p_2} a_2^\dagger a_2 +({\omega_{p_1}-\omega_{p_2}})b_a^\dagger b_a $, the Hamiltonian in \autoref{eq:eq1} becomes
\begin{eqnarray}\label{eq:eq3}
H'&=- \Delta_1 a_1^\dagger  a_1 + \Delta_a b_a^\dagger b_a+\omega_m b_m^\dagger b_m- g_{m} a_1^\dagger a_1 (b_m +b_m^\dagger)\nonumber \\&+{J_m}({e^{i\theta}}{b_{a}^{\dagger}}{b_m} + {e^{-i\theta}}{b_a}{b_{m}^{\dagger}})+ \frac{\eta}{2}(b_m + b_m^\dagger)^4\nonumber \\& +iE_1(a_1^{\dagger } - a_1) - G_a( a_1^\dagger b_a+ a_1 b_a^\dagger),
\end{eqnarray}
where we have defined $\Delta_1=\omega_{p_1}-\omega_{c_1}$, and $\Delta_a=\omega_a+\omega_{p_2}-\omega_{p_1}$. Here $\alpha_2$ is the steady-state of the control optical mode $a_2$, which has been treated classically as it is assumed to be strong compared to the weak strength of the Brillouin acoustic mode $b_a$ (see details in  \cref{App.A}).

\begin{figure}[tbh]
  \begin{center}
  \resizebox{0.4\textwidth}{!}{
  \includegraphics{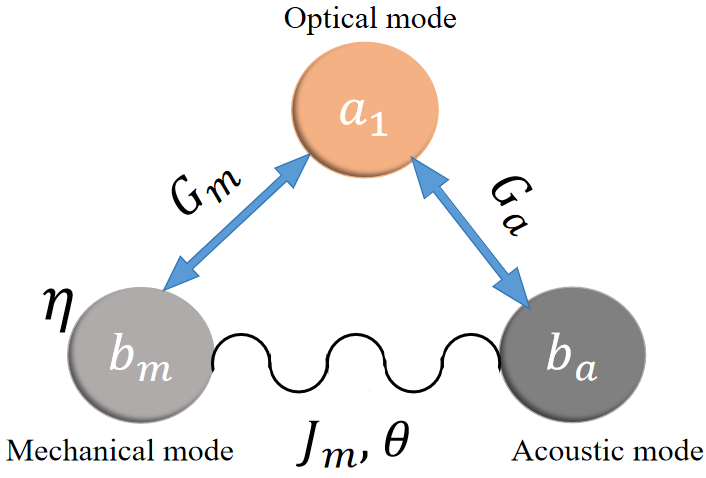}}
  \end{center}
\caption{Sketch of our linearized three mode optomechanical system. The mode $a_1$ is coupled to the acoustic ($b_a$) and the mechanical mode ($b_m$) with coupling strength $G_a$ and $G_m$ respectively. The mechanical oscillator has the Duffing coefficient $\eta$. The phonon-phonon hopping rate $J_m$ is modulated by the phase $\theta$.}
\label{fig:Fig1}
\end{figure}

By following the standard linearization procedure (\cref{App.A}), the Hamiltonian in \autoref{eq:eq3} can be linearized, and that reduces our scheme to a three-mode optomechanical system as depicted in \autoref{fig:Fig1}. In order to investigate the squeezing of the mechanical resonator $b_m$, we introduce the following Bogoliubov transformation,
\begin{equation}\label{eq:Bogoliubov}
b_m=b_m^s \cosh(r)-b_m^{s\dagger} \sinh(r),
\end{equation}
 with the defined squeezing parameter $r=\frac{1}{4}\ln(1+\frac{2\Lambda}{\omega_m})$ and the squeezed mechanical mode $b_m^s$. This transformation leads to the new linearized Hamiltonian, 
\begin{eqnarray}\label{eq:linear}
&H_{lin}^s=-\tilde{\Delta} \delta a_1^\dagger  \delta a_1 +  \Delta_a \delta b_a^\dagger \delta b_a+\omega'_m \delta b_m^{s\dagger}\delta b_m^s  \nonumber \\&- (G'_m\delta a_1^\dagger+G_m^{'\ast} \delta a_1) (\delta b_m^{s\dagger} +\delta b_m^{s})+{J'_m}({e^{i\theta}}{\delta b_{a}^{\dagger}}\delta b_m^s \nonumber \\&+ {e^{-i\theta}}{\delta b_a}\delta b_m^{s\dagger})  - {G_a} ( \delta a_1^\dagger \delta b_a + \delta a_1 \delta b_a^\dagger ),
\end{eqnarray}
with the effective parameters $\omega'_m= \sqrt{\omega_m{(\omega_m +2\Lambda)}}$, $G'_m=G_me^{-r}$ and $J'_m=J_m \cosh(r)$. The effective mechanical coupling is $G_m=g_m\alpha_1$, where $\Lambda=24\eta\Re(\beta_m)^2$. $\beta_m$ ($\alpha_1$) is the steady-state of the mechanical (optical) mode. 

By taking into account the dissipation associated with the optical ($\kappa$), the mechanical ($\gamma_m$) and the acoustic ($\gamma_a$) modes, one can obtain the following equations describing the dynamics of the fluctuations operators,  
\begin{equation}\label{eq:fluc}
\begin{cases}
\delta\dot{a}_1 &= \left(i\tilde{\Delta} - \frac{\kappa}{2}\right) \delta a_1 + iG'_m({\delta}b_{m}^{s\dagger} +{\delta}b_m^s) +i{G_a} \delta b_a \\&+\sqrt{\kappa}a_1^{in}\\ 
\delta\dot{b}_a &= - ({\frac{\gamma_a}{2}} + i{\Delta_a})\delta{b}_a  - i{J'_{m}}e^{i\theta}\delta{b}_m^s +i{G_a} \delta a_1 \\&+\sqrt{\gamma_a}b_{a}^{in} \\
\delta\dot{b}_m^s &= -({\frac{\gamma_m}{2}}+ i {\omega'_m})\delta{b}_m^s   -i{J'_{m}}e^{-i\theta}\delta{b}_a + i(G_m^{'\ast}\delta a_1 \\&+ G_m'\delta a_1^\dagger ) + \sqrt{\gamma_m}b_{m}^{s^{in}}, 
\end{cases}
\end{equation}
where $\tilde{\Delta}=\Delta-2g_m\rm Re(\beta_m)$ is the effective detuning. In the set of \autoref{eq:fluc}, $a_1^{in}$, $b_a^{in}$ and $b_m^{s^{in}}$ are zero-mean noise operators characterized by the following auto-correlation functions,
\begin{eqnarray}\label{eq:noise}
&\langle a_1^{in}(t)a_1^{in\dagger}(t') \rangle &=\delta(t-t'), \hspace*{0.5cm} \langle a_1^{in\dagger}(t)a_1^{in}(t') \rangle = 0, \nonumber \\
&\langle b_a^{in}(t)b_a^{in\dagger}(t') \rangle &= \delta(t-t'), \hspace*{0.5cm} \langle b_a^{in\dagger}(t)b_a^{in}(t') \rangle = 0 , \nonumber \\
&\langle b_m^{s^{in}}(t)b_m^{s^{in}\dagger}(t') \rangle &= (n_{th}\cosh(2r)+\sinh^2(r))\delta(t-t'), \nonumber \\ 
&\langle b_m^{in}(t)b_m^{in}(t') \rangle &= (n_{th}+\frac{1}{2})\sinh(2r)\delta(t-t').
\end{eqnarray}

Here, $n_{th}$ represents the equilibrium phonon occupation number of the mechanical resonator, and is defined as, $n_{th}=[\rm exp(\frac{\hbar \omega_m}{k_bT})-1]^{-1}$, where $\rm k_b$ is the Boltzmann constant. Owing to the high-frequency acoustic mode $b_a$ compared to the mechanical one ($\omega_m \ll \omega_a$) the thermal acoustic phonon occupation has been neglected. In order to investigate the effect of the mechanical coupling $J_m$ and its phase modulation $\theta$ on squeezing of the targeted mechanical resonator, we first define the following amplitude (position) and phase (momentum) quadrature operators: $\delta X_{\mathcal{O}} =\frac{\delta \mathcal{O}^\dagger + \delta \mathcal{O} }{\sqrt{2}}$, $\delta Y_{\mathcal{O}} =i\frac{\delta \mathcal{O}^\dagger - \delta \mathcal{O}}{\sqrt{2}}$, with $\mathcal{O}\equiv a_1,b_a,b_m^s$. Similarly, the related noise quadratures read as: $\delta X_{\mathcal{O}}^{in} =\frac{\delta \mathcal{O}^{\dagger in} + \delta \mathcal{O}^{in}}{\sqrt{2}}$, $\delta Y_{\mathcal{O}}^{in} =i\frac{\delta \mathcal{O}^{\dagger in} - \delta \mathcal{O}^{in}}{\sqrt{2}}$. This enables us to derive the set of equations describing the quadrature dynamics of our system as, 
\begin{equation}\label{eq:quadra}
\dot{u}={\rm M} u+z^{in}.
\end{equation}
Here ${u}=(\delta X_{a_1},\delta Y_{a_1}, \delta X_{b_a},\delta Y_{b_a}, \delta X_{b_m},\delta Y_{b_m})^T$, $z^{in}=(\sqrt{\kappa} X_{a_1}^{in},\sqrt{\kappa} Y_{a_1}^{in},\sqrt{\gamma_a} X_{b_a}^{in},\sqrt{\gamma_a} Y_{b_a}^{in},\sqrt{\gamma_m} X_{b_m}^{in},\sqrt{\gamma_m} Y_{b_m}^{in},)^T$ and the matrix $\rm M$ is given by,
\begin{equation}\label{eq:matrix}
{\rm M}=
\begin{pmatrix}
-\frac{\kappa_1}{2}&-\tilde{\Delta}&0&-G_a&0&0 \\
\tilde{\Delta}&-\frac{\kappa_1}{2}&G_a&0&2G_m^{'}&0 \\ 
0&-G_a&-\frac{\gamma_a}{2}&\Delta_a & J'_m\sin{\theta} & J'_m \cos{\theta} \\
G_a&0&-\Delta_a &-\frac{\gamma_a}{2}&-J'_m \cos{\theta} & J'_m\sin{\theta} \\
0&0&-J'_m\sin{\theta}&J'_m\cos{\theta}&-\frac{\gamma_m}{2}&\omega_m^{'} \\
2G_m^{'}&0&-J'_m \cos{\theta}&-J'_m\sin{\theta}&-\omega_m^{'}& -\frac{\gamma_m}{2}.
\end{pmatrix},
\end{equation}
Here the effective couplings $G_m^{'}$ and $G_a$ have been assumed to be real for simplicity.

\section{Synthetic magnetism enhanced mechanical squeezing}\label{sec:squeez}
To illustrate the enhancement of the mechanical squeezing in our proposed scheme, we need to derive the position variance of the nonlinear mechanical resonator. For this purpose, we evaluate the covariance matrix, $V_{ij}$, where $V_{ij}=\frac{\langle u_i u_j  + u_j u_i \rangle}{2}$. The covariance matrix satisfies the equation of motion,
\begin{equation}\label{eq:lyap1}
\dot{V}={\rm M}V+V{\rm M^T}+D,
\end{equation}
with the diagonal diffusion matrix $D=Diag[\frac{\kappa}{2},  \frac{\kappa}{2}, \frac{\gamma_a}{2}, \frac{\gamma_a}{2},\\ \frac{\gamma_m}{2}e^{2r}(2n_{th} + 1), \frac{\gamma_m}{2}e^{-2r}(2n_{th} + 1)]$, and the matrix $\rm M$ must meet the Routh-Hurwitz stability criterion where all its eigenvalues should have negative real parts. The parameters are chosen in such a way that this stability criterion is met and they are experimentally feasible.  As we aim to study the steady-state behavior of the mechanical squeezing, we use the long time limit of \autoref{eq:lyap1} which is known as the Lyapunov equation,

\begin{equation}\label{eq:lyap}
{\rm M}V+V{\rm M^T}=-D.
\end{equation}

\begin{figure}[tbh]
\begin{center}
  \resizebox{0.5\textwidth}{!}{
  \includegraphics{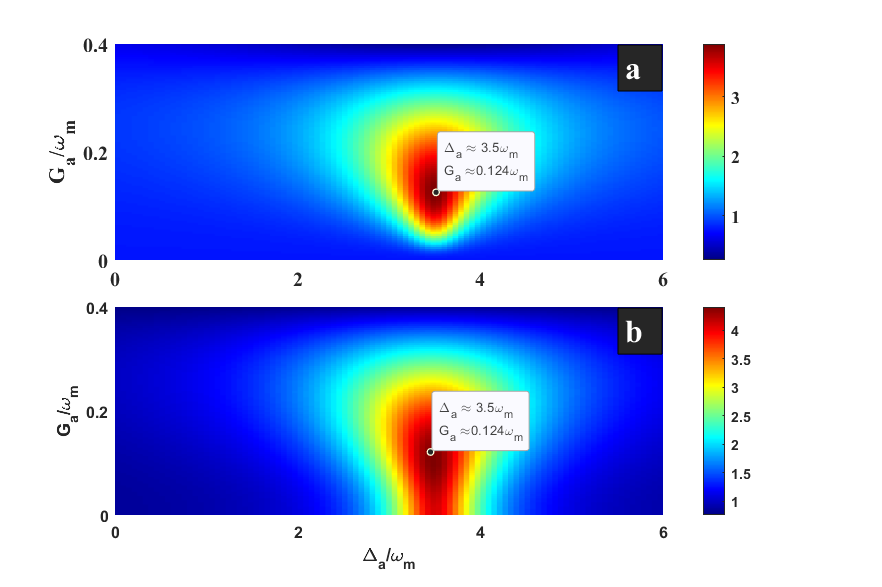}}
  \end{center}
\caption{Mechanical position variance from \autoref{eq:VardB} versus $G_a$ and $\Delta_a$. The mechanical coupling strength is $J_m=0$ for (a) and $J_m=0.1\omega_m$ for (b).  The parameters used are $\omega_m/2\pi=1\rm {MHz}$, $g_m=10^{-4}\omega_m$, $\kappa=0.02\omega_m$, $\gamma_a=0.4\omega_m$, $\gamma_m=10^{-4}\omega_m$, $\eta=10^{-4}\omega_m$, $n_{th}=100$, $G_m=0.15\omega_m$, $\tilde{\Delta}=-\omega'_m$, and $\theta=\pi/2$.}
\label{fig:Fig2}
\end{figure}

From the $V_{ij}$ elements, we can obtain the position variance ($\langle \delta q_m^2 \rangle$) of the mechanical resonator as follows:
\begin{equation}\label{var}
\langle \delta q_m^2 \rangle = V_{55} e^{-2r},
\end{equation}
which can be expressed in \rm{dB} units as,
\begin{equation}\label{eq:VardB}
\langle \delta q_m^2 \rangle(dB) = -10\log_{10}\frac{\langle \delta q_m^2 \rangle}{\langle \delta q_{ZPF}^2 \rangle},
\end{equation}
where
$\langle \delta q_{ZPF}^2 \rangle=\frac{1}{2}$ is the zero-point fluctuations of the mechanical resonator. One can clearly see that the position quadrature is getting squeezed. In order to effectively suppress the thermal effects at the the optimal acoustic resonance $\Delta_a^{\rm {opt}}$, as depicted in the terms, $G_m^{'}(\delta a_1^{\dagger}\delta b_m^{s\dagger}+\delta a_1\delta b_m^s)$, we work in the red-detuned regime ($\tilde\Delta=-\omega_m^{'}$). The main parameters used for our numerical simulations are $\omega_m/2\pi=1\rm {MHz}$, $g_m=10^{-4}\omega_m$, $\kappa=0.02\omega_m$, $\gamma_a=0.4\omega_m$, $\gamma_m=10^{-4}\omega_m$, $\eta=10^{-4}\omega_m$, $n_{th}=100$, $G_m=0.15\omega_m$, and $\tilde{\Delta}=-\omega'_m$. The other parameters such as $J_m$, $\theta$, $G_a$ and $\Delta_a$ are adjusted as latter on indicated on the related figures.   \autoref{fig:Fig2} exhibits the position variance as a function of the the acoustic effective coupling $G_a$ and detuning $\Delta_a$. It can be observed that the largest squeezing is generated at the acoustic resonance  $\Delta_a^{\rm {opt}}\approx3.5\omega_m$ and $G_a=0.124\omega_m$. For the rest of the work, we take $\Delta_a=\Delta_a^{\rm {opt}}$ and assume the red-sideband detuning resonance condition. It is revealed in Fig. 2(a) that no squeezing is generated near $G_a\sim0$, or it is very weak. However, Fig.2(b) shows a large degree of squeezing generated even for $G_a=0$ around the optimal acoustic resonance $\Delta_a^{\rm {opt}}$. This feature reveals how the synthetic gauge induces the squeezing even if the BSBS effect is not supported in the system. Moreover, it could be seen that the squeezing generated in the presence of $J_m\neq0$ is stronger compared to the case when there is no mechanical coupling $J_m=0$. To get further insight into the enhancement of the squeezing through the synthetic gauge, we study the mechanical position variance against the variation of both $J_m$ and $\theta$, as displayed in Fig.3(a). On the other hand, corresponding study for the effective phonon number is depicted in Fig.3(b). It should be noted that the effective phonon number corresponding to the mechanical oscillator $n_{eff}^m$ can also be expressed in term of the $V_{ij}$ elements, and it yields,
\begin{eqnarray}\label{eq:num1}
n_{eff}^m= \frac{1}{2}\left(V_{55}e^{-2r}  + V_{66}e^{2r} - 1 \right). 
\end{eqnarray}    
\begin{figure}[tbh]
\begin{center}
  \resizebox{0.5\textwidth}{!}{
  \includegraphics{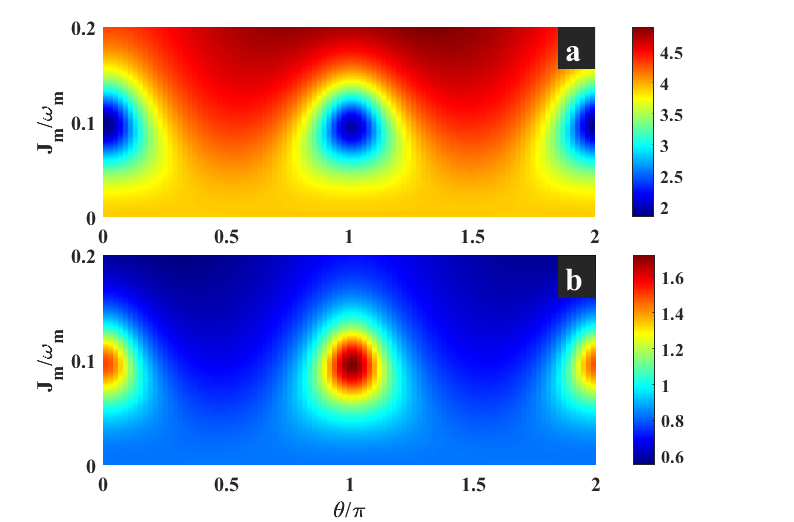}}
  \end{center}
\caption{(a) Mechanical position variance (\autoref{eq:VardB}) and (b) effective (mechanical) phonon number (\autoref{eq:num1}) versus $J_m$ and $\theta$. Optimal squeezing parameters deduced from \autoref{fig:Fig2} are used: $G_m=0.15\omega_m$, $G_a=0.124\omega_m$, and $\Delta_a^{\rm {opt}}$. The others parameters are similar to the ones in \autoref{fig:Fig2}.}
\label{fig:Fig3}
\end{figure}
It can be seen from \autoref{fig:Fig3}(a,b) that both variance and phonon number are modulated along the $\theta$ direction. For $J_m\le 5\times10^{-2}\omega_m$, the effect of the phase $\theta$ is negligible. However, for $J_m > 5\times10^{-2}\omega_m$, the variance (\autoref{fig:Fig3}a) exhibits peaks at $\theta\equiv(n+\frac{1}{2})\pi$ and dips for $\theta\equiv n\pi$, $n$ being an integer. This feature shows that significant squeezing is generated at $\theta\equiv(n+\frac{1}{2})\pi$. This also corresponds to an effective minimum phonon number as expected in \autoref{fig:Fig3}b (see near $J_m\sim0.1\omega_m$ for instance). By paying attention to the colorbars of these figures, it can be figured out that with  an increase in mechanical coupling $J_m$, the mechanical resonator is getting cooled down significantly resulting in stronger degree of squeezing. These figures point out the crucial role played by both $J_m$ and $\theta$ regarding enhancement of squeezing in our proposal. To further reveal the oscillatory feature of the squeezing above discussed, we have extracted the position variance and phonon number from \autoref{fig:Fig3} at $J_m\sim0.1\omega_m$, and that is displayed in \cref{App.B}.       

\begin{figure*}[tbh]
  \begin{center}
  \resizebox{1.0\textwidth}{!}{
  \includegraphics{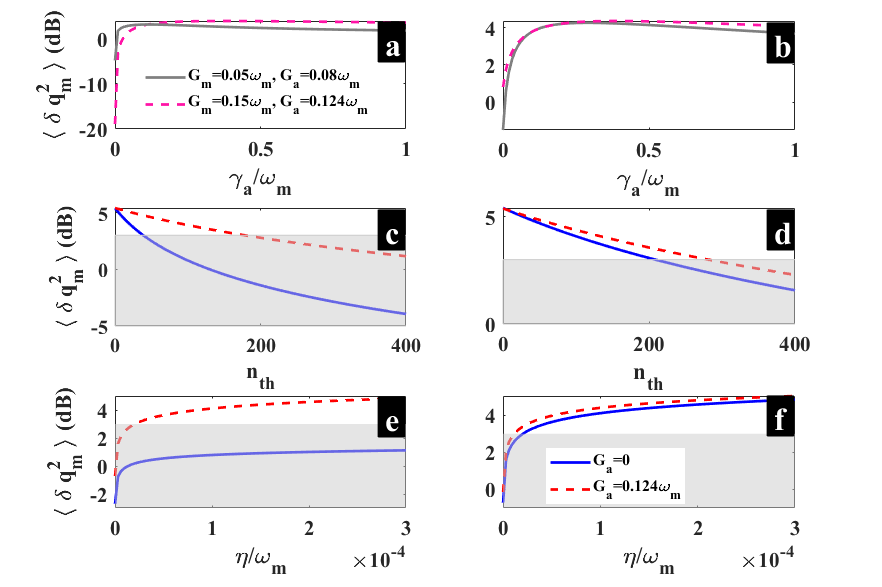}}
  \end{center}
\caption{Left and right column depicts the same quantities, which are plotted for $J_m=0$ and $J_m=0.1\omega_m$, respectively. (a,b) display the position variance of the mechanical resonator over the Brillouin acoustic decay rate $\gamma_a$  for two specific couples of points ($G_a, G_m$): ($0.05\omega_m, 0.08\omega_m$) for the full line, and ($0.15\omega_m, 0.124\omega_m$) for the dashed line. (c,d) analyze the robustness of the mechanical position variance versus the thermal noise $n_{th}$. (e,f) figure out the behavior of the mechanical position variance versus the Duffing coefficient $\eta$. In (c-f), the full line is for $G_a=0$ and the dashed line is for $G_a=0.124\omega_m$. The rest of the parameters are the same as in \autoref{fig:Fig2}.}
\label{fig:Fig4}
\end{figure*}

Next, we investigate the role of synthetic gauge in enhancing the squeezing of the position quadrature. We take specific parameters in the ($G_m, G_a$) space, and in Fig. 4 depict how the position variance varies with $\gamma_a$,$n_{th}$ and $\eta$. The left and right columns in \autoref{fig:Fig4} depict the mechanical position variance plotted over the same variable for $J_m=0$ and $J_m=0.1\omega_m$, respectively. The first interesting parameter considered here is the Brillouin acoustic decay rate $\gamma_a$. The BSBS process involved in our proposal relies on the fact that $\gamma_a\gg \kappa(\gg\gamma_m)$. Therefore, it is crucial to point out the impact of this condition on the squeezing generation as shown in \autoref{fig:Fig4}(a,b). Without the mechanical coupling ($J_m=0$), \autoref{fig:Fig4}a shows how the variance increases rapidly and reaches a saturation limit above $\gamma_a\gg \kappa=2\times10^{-2}\omega_m$ as expected. Above this BSBS condition, the position variance settles on a sort of plateau, where the acoustic decay rate does no longer affect the behavior of the generated squeezing. This regime could be of great interests for quantum technologies involving squeezing such as quantum information processing, quantum sensing and metrology, and quantum computing. By considering the synthetic gauge ($J_m\neq0$), \autoref{fig:Fig4}b shows a great enhancement of the degree of squeezing (from $0$ to near $\rm{5dB}$) compared to when $J_m=0$ (compare \autoref{fig:Fig4}a to \autoref{fig:Fig4}b). Despite the fact that the position variance behaves with almost the same shape in both \autoref{fig:Fig4}(a,b), it is worth to mention that the merit of the synthetic gauge has been to push the limit of the generated squeezing beyond the $\rm{3dB}$. Furthermore, these figures show that the more the couplings ($G_m, G_a$) are enhanced, better is the degree of the squeezing (see \cref{App.B} for a large view). In \autoref{fig:Fig4}c, we have plotted the position variance over the thermal phonon mechanical excitation. The shadow area depicts the region below the $\rm{3dB}$. It can be seen that when $G_a=0$, there is only a short window where the squeezing is above the $\rm{3dB}$ (full line) compared to when $G_a\neq0$ (dashed line). This feature reveals the key role played by the BSBS effect, which suppresses the heating processes in the system \cite{BSBS_basis}. This BSBS effect is further reinforced through the gap between the two lines in \autoref{fig:Fig4}c, showing how $G_a\neq0$ has improved the degree of squeezing compared to the case $G_a=0$. By taking into account the synthetic gauge, we observe in \autoref{fig:Fig4}d that the gap between the cases $G_a=0$ and $G_a\neq0$ has been efficiently reduced. Indeed, the squeezing generated for $G_a=0$ follows almost the one generated for $G_a\neq0$, and they stay longer beyond the $\rm{3dB}$ compared to what is shown in \autoref{fig:Fig4}d. This confirms that the synthetic gauge contributes to suppress the heating channels, inducing strong squeezing even for $G_a=0$, as earlier discussed in \autoref{fig:Fig2}. Furthermore, it can be observed that squeezing is somehow less prone to the affect of thermal phonons in the presence of synthetic gauge. In \autoref{fig:Fig4}(e,f), we display the mechanical position variance versus the Duffing nonlinear coefficient $\eta$. These figures show how the variance sharply increases for weak values of $\eta$, and settles quickly to a plateau like a saturation limit. However, it can be observed that the synthetic gauge bridges the gap between the cases $G_a=0$ and $G_a\neq0$ as previously discussed. Moreover, \autoref{fig:Fig4}e shows how the degree of the generated squeezing for $G_a=0$ is below the $\rm{3dB}$, while it exceeds this limit for $J_m\neq0$  as depicted in \autoref{fig:Fig4}f. Once again, this highlights the merit of the synthetic gauge that induces a strong squeezing even for $G_a=0$ in our proposal as shown in \autoref{fig:Fig2}.  

\begin{figure}[tbh]
\begin{center}
  \resizebox{0.5\textwidth}{!}{
  \includegraphics{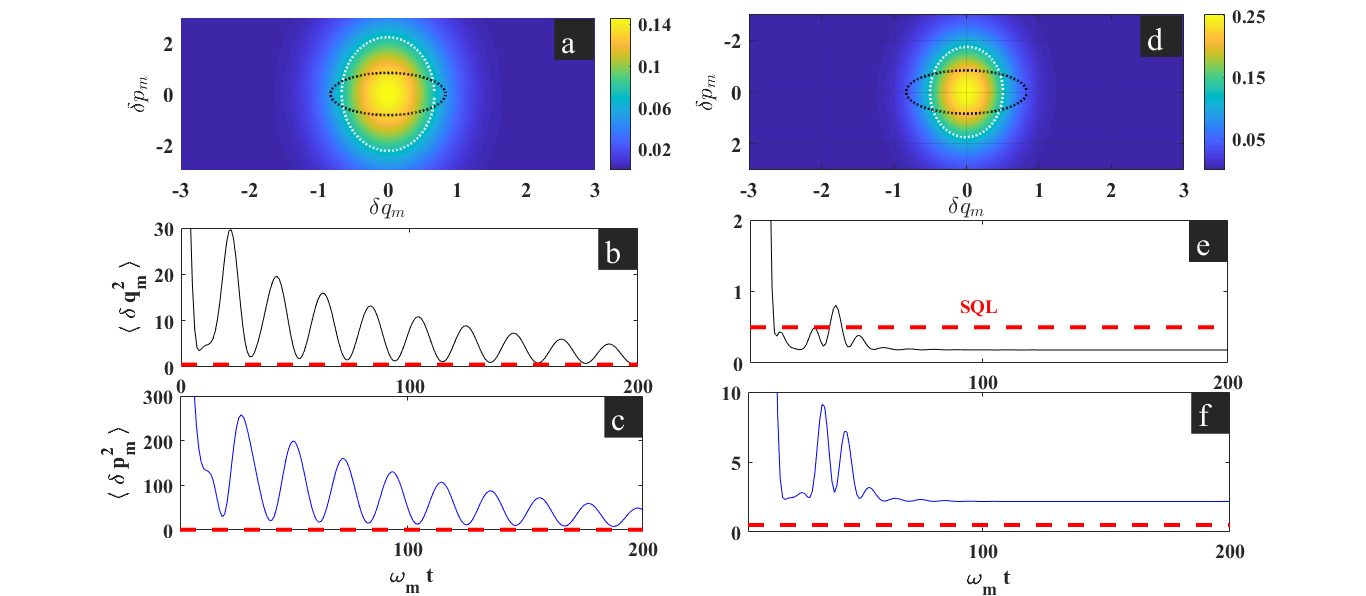}}
  \end{center}
\caption{Wigner function distributions and dynamic of mechanical variances. Left column depicts Wigner function (a), position variance (b) and the momentum variance (c) at $J_m=0.1\omega_m$, and $\theta=0$. Right column depicts Wigner function (d), position variance (e) and the momentum variance (f) at $J_m=0.1\omega_m$, and $\theta=\pi/2$. These figures are extracted from the feature displayed in \autoref{fig:Fig3}.  The other parameters are the same as in \autoref{fig:Fig2}.}
\label{fig:Fig5}
\end{figure}

In Fig. 5(a) and (d) we exhibit the Wigner function distribution for the squeezed state generated for some specific parameters. Owing to the Gaussian nature of the quantum noise, our linearized system can be described through a single Gaussian Wigner function in the steady state defined as \cite{Banerjee_2023}
\begin{equation}\label{eq:wigner}
W(u_m)=\rm{\frac{1}{2\pi\sqrt{det[V_m]}}\exp\big[-\frac{u_m^TV_m^{-1}u_m}{2}\big]},
\end{equation}
where $u_m=(\delta q_m, \delta p_m)^T$ is the column vector of the mechanical fluctuations and $V_m$ stands for the covariance matrix for the mechanical mode. The left column in \autoref{fig:Fig5} depicts the Wigner function distribution of the mechanical resonator (\autoref{fig:Fig5}a), the mechanical position variance (\autoref{fig:Fig5}b) and the mechanical momentum variance (\autoref{fig:Fig5}c) for the synthetic modulation phase of $\theta=0$. Similarly, the right column displays the same quantities at $\theta=\frac{\pi}{2}$ i.e., the Wigner function distribution (\autoref{fig:Fig5}d), the position variance (\autoref{fig:Fig5}e) and the momentum variance (\autoref{fig:Fig5}f). These figures have been plotted in the long time limit so that the system has reached its steady state. Moreover, these figures have been extracted from \autoref{fig:Fig3}a for the mechanical coupling strength fixed at $J_m=0.1\omega_m$. In \autoref{fig:Fig5}(a,d), the contour represented by a dark circle depicts the related coherent state, while the shrunken and expanded shape bounded by the white ellipse features the corresponding generated squeezed state. As predicted from \autoref{fig:Fig3}a, the Wigner function at $\theta=0$ (\autoref{fig:Fig5}a) shows less squeezing as compared to when $\theta=\pi/2$ (\autoref{fig:Fig5}d). Moreover, these Wigner function distributions confirm the squeezing along the position direction, while the quantum fluctuation is transferred to the momentum direction which is amplified. To further analyze these observations, we have plotted the dynamical time-evolution of the involved variances. As expected, both position and momentum dynamical evolution at $\theta=0$ clearly show no squeezing (\autoref{fig:Fig5}(b,c)), which is revealed by an oscillatory behavior above the Standard Quantum Limit (\rm SQL). However, we can qualitatively observe that there is less amount of noise along the position direction compared to the momentum direction as predicted by the corresponding Wigner distribution (\autoref{fig:Fig5}a). For $\theta=\pi/2$, however, \autoref{fig:Fig5}e shows squeezing in position quadrature below the \rm{SQL}, while its corresponding momentum is far above the \rm{SQL}, revealing that quantum fluctuation has been amplified along this direction as predicted from \autoref{fig:Fig5}d. From the above analysis, it can be seen that our Wigner function distribution together with the dynamical evolution of the variances agree well with the synthetic magnetism squeezing enhancement pointed out in \autoref{fig:Fig3}. This work provides an efficient scheme towards enhancement of squeezing beyond $\rm{3dB}$ in optomechanical system that is based on the backward stimulated Brillouin scattering effect. The generated squeezing under this scheme is robust enough against thermal noise compared to the case without the synthetic magnetism. This investigation can be extended to electromechanical systems and hybrid opto-electromechanical systems.        

\section{Conclusion} \label{sec:concl}
We have investigated the synthetic magnetism effect inducing mechanical squeezing enhancement in an optomechanical system, which hosts a backward stimulated Brillouin scattering process. Our model system consists of an acoustic mode that couples to two optical modes through the BSBS process, and a nonlinear mechanical oscillator that couples to the two optical modes through the standard optomechanical radiation pressure. A mechanical coupling, with a strength $J_m$ that is modulated through a phase $\theta$, is connecting the acoustic and mechanical modes, which induces a synthetic magnetism in our proposal. Without this synthetic magnetism, there is a given amount of squeezing that is generated in the system. This squeezing is mainly induced by the BSBS process, and is fragile against thermal noise. When the synthetic magnetism is accounted, the degree of the generated squeezing is greatly enhanced and goes far beyond the \rm{3dB}. Moreover, this large induced squeezing persists even when the system is free from the BSBS process. Furthermore, this generated squeezing is robust enough against thermal noise compared to the case without synthetic magnetism. Another merit of the synthetic magnetism in our proposal is revealed through the peaks and dips of both squeezing magnitude and mechanical effective phonon number depending on the modulation phase of the mechanical coupling. This oscillatory feature is reminiscent of a sudden death and revival of squeezing phenomenon, which can be used to maintain a desired magnitude of squeezing by tuning the phase $\theta$. Our work sheds light on a flexible scheme that can be used to generate a large amount of mechanical squeezing, far beyond the $\rm{3dB}$ limit. Our scheme can be implemented in optical and microwaves cavities, as well as in hybrid optomechanical systems. Such a generated squeezed states can be useful for a range of quantum applications including quantum information processing, quantum sensing and metrology, and the recent development in quantum computing.

\section*{Acknowledgments}
This work has been carried out under the Iso-Lomso Fellowship at Stellenbosch Institute for Advanced Study (STIAS), Wallenberg Research Centre at Stellenbosch University, Stellenbosch 7600, South Africa. S. Agasti wishes to acknowledge the European Union H2020 MSCA; Project number: 101065991 (acronym: SingletSQL) for supporting the work.  A.K. Sarma acknowledges the STARS scheme, MoE, government of India (Proposal ID 2023-0161). K.S. Nisar is grateful to the funding from Prince Sattam bin Abdulaziz University, Saudi Arabia project number (PSAU/2024/R/1445). The authors are thankful to the Deanship of Graduate Studies and Scientific Research at University of Bisha for supporting this work through the Fast-Track Research Support Program.




\appendix

\section{Effective Hamiltonian}\label{App.A}
The Hamiltonian that describes the system considered by us is given by,
\begin{align}
H&=\sum_{j=1,2} \omega_{c_j} a_j^\dagger  a_j + \omega_a b_a^\dagger b_a + \omega_m b_m^\dagger b_m - \sum_{j=1,2}g_{m,a}a_j^\dagger a_j   (b_m +b_m^\dagger)\\ \nonumber &+ \frac{\eta}{2}(b_m + b_m^\dagger)^4 +J_m({e^{i\theta}}{b_{a}^{\dagger}}{b_m} + {e^{-i\theta}}{b_a}{b_{m}^{\dagger}})\\ \nonumber&+\sum_{j=1,2}iE_j(a_j^{\dagger }e^{-i\omega_{p_j} t} - a_je^{i\omega_{p_j} t})- g_a( a_1^\dagger a_2 b_a+ a_1 a_2^\dagger b_a^\dagger).
\end{align}

The total Hamiltonian of the hybrid system in a frame rotating with the laser frequency $\omega_{p_j}$ is given by: 

\begin{align}\label{eq:hamil}
H'&=-\sum_{j=1,2} \Delta_{j} a_j^\dagger  a_j +  \Delta_a b_a^\dagger b_a+\omega_m b_m^\dagger b_m- \sum_{j=1,2}g_{m,a}a_j^\dagger a_j (b_m +b_m^\dagger)\\ \nonumber &+ \frac{\eta}{2}(b_m + b_m^\dagger)^4 +J_m({e^{i\theta}}{b_{a}^{\dagger}}{b_m} + {e^{-i\theta}}{b_a}{b_{m}^{\dagger}})\\ \nonumber &+iE_1(a_1^{\dagger } - a_1) +iE_2(a_2^{\dagger} - a_2) - g_a( a_1^\dagger a_2 b_a+ a_1 a_2^\dagger b_a^\dagger),
\end{align}
with $\Delta_j=\omega_{p_j}-\omega_{c_j}$, and $\Delta_a=\omega_a+\omega_{p_2}-\omega_{p_1}$. 
By considering that the control field $a_2$ is strong enough compared to $a_1$, it can be treated classically by deriving its steady-state as,
\begin{align}
\dot{a}_2 &= i[H',a_2], \\
 &=(i\Delta_2-\frac{\kappa_2}{2})a_2 + ig_{m}a_2(b_m+b_m^\dagger) + E_2 + ig_a a_1 b_a^\dagger \nonumber \\
 &=(i\Delta'_2-\frac{\kappa_2}{2})a_2 + E_2 + ig_a a_1 b_a^\dagger,
\end{align}
with $\Delta'_2=\Delta_2 + g_{m}a_2(b_m+b_m^\dagger)$. The steady-state solution ($\dot{a}_2=0$) yields, 
\begin{align}
\alpha_{2}&\sim\frac{-E_2}{i\Delta'_2-\frac{\kappa_2}{2}} \hspace{1em} \text{or} \hspace{1em} |\alpha_{2}|\sim\frac{E_2}{\sqrt{\Delta_2^{'^2}+\frac{\kappa_2^2}{4}}}.
\end{align}
By substituting this expression in the rest of the Hamiltonian in \autoref{eq:hamil}, we get the following reduced Hamiltonian,
\begin{align}
H'&=-\Delta_{1} a_1^\dagger  a_1 +  \Delta_a b_a^\dagger b_a+\omega_m b_m^\dagger b_m- g_{m} a_1^\dagger a_1 (b_m +b_m^\dagger)\nonumber \\ & + \frac{\eta}{2}(b_m + b_m^\dagger)^4 +J_m({e^{i\theta}}{b_{a}^{\dagger}}{b_m} + {e^{-i\theta}}{b_a}{b_{m}^{\dagger}})\\ \nonumber & +iE_1(a_1^{\dagger } - a_1)- {G_a} ( a_1^\dagger b_a + a_1 b_a^\dagger ),
\end{align}
where the acoustic effective coupling is $G_a=g_a\alpha_{2}$ as mentionned in the main text.

\section{Oscillatory behavior and effect of the effective couplings} \label{App.B}
\begin{figure}[tbh]
  \begin{center}
  \resizebox{0.45\textwidth}{!}{
  \includegraphics{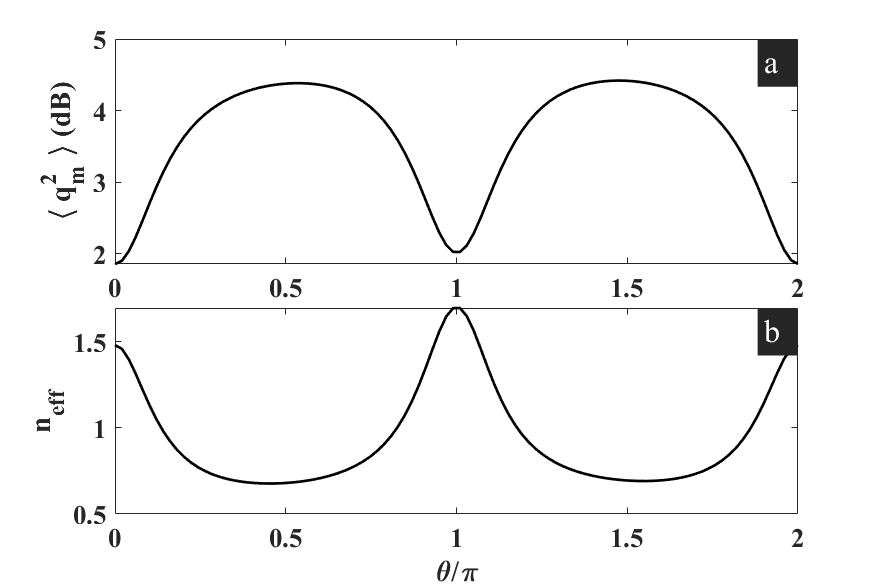}}
  \end{center}
\caption{Mechanical position variance (a) and effective phonon number (b) (extracted from \autoref{fig:Fig3} at $J_m=0.1\omega_m$) against variation of phase $\theta$. The other parameters are the same as in \autoref{fig:Fig2}.}
\label{fig:FigA1}
\end{figure}

This section highlights some phenomena aforementioned in the main text related to \autoref{fig:Fig3} and \autoref{fig:Fig4}. Indeed, it was mentioned that the position variance and the phonon number displayed in \autoref{fig:Fig3} exhibit sort of oscillatory behavior depending on the synthetic phase $\theta$ around the phonon hopping rate $J_m=0.1\omega_m$. Moreover, it was pointed out that these two quantities are out of phase with each other over $\theta$. \autoref{fig:FigA1}a depicts the position variance while \autoref{fig:FigA1}b shows variation of effective phonon numbers against phase $\theta$. These depictions are extracted from  \autoref{fig:Fig3}a and \autoref{fig:Fig3}b at $J_m=0.1\omega_m$,  respectively. As expected, it can be clearly seen how these two quantities exhibit oscillatory features and are out of phase.    

\begin{figure}[tbh]
  \begin{center}
  \resizebox{0.45\textwidth}{!}{
  \includegraphics{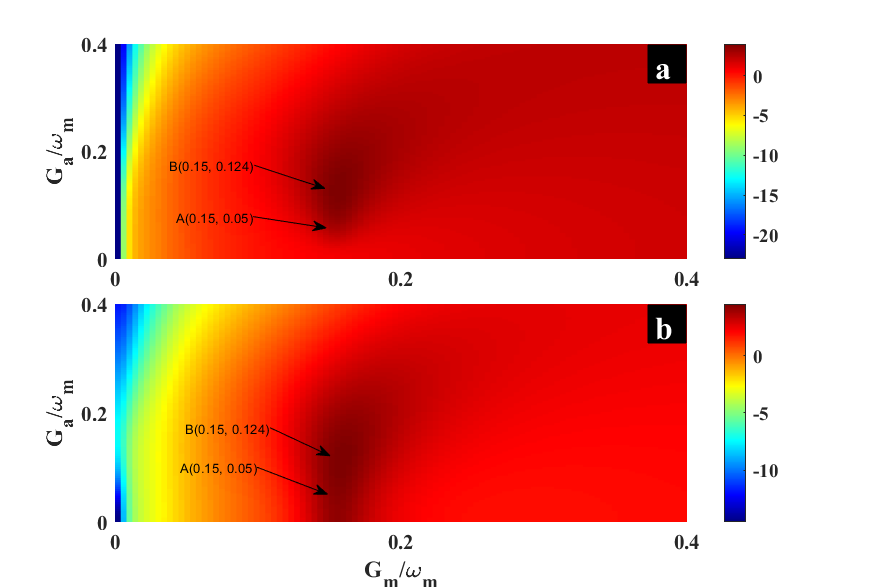}}
  \end{center}
\caption{Squeezing degree of the mechanical position variance in $G_a$ and $G_m$ space for $J_m=0$ (a) and $J_m=0.1\omega_m$ (b). The rest of the parameters are the same used in \autoref{fig:Fig2}.}
\label{fig:FigA2}
\end{figure}

In \autoref{fig:Fig4}, it has been pointed out that as the effective coupling $G_a$ and $G_m$ are enhanced, the strength of squeezing gets enhanced as well. To get a clearer view of that feature, \autoref{fig:FigA2} displays the mechanical position variance in $G_a$ and $G_m$ space for $J_m=0$ (a) and $J_m=0.1\omega_m$ (b). As it can be seen in both figures, the squeezing get stronger and stronger as the strength of the two effective couplings increases. This justifies our statement aforementioned in the main text. Moreover, by paying attention to the colorbars in \autoref{fig:FigA2}, it appears that the squeezing is enhanced as the synthetic magnetism is accounted in the system, i.e., $J_m\neq0$ and $\theta=\pi/2$. This can be seen by comparing the couple of points highlighted in \autoref{fig:FigA2} for instance.    

\newpage

\bibliography{Squeezing}

\begin{thebibliography}{45}
\expandafter\ifx\csname natexlab\endcsname\relax\def\natexlab#1{#1}\fi
\expandafter\ifx\csname bibnamefont\endcsname\relax
  \def\bibnamefont#1{#1}\fi
\expandafter\ifx\csname bibfnamefont\endcsname\relax
  \def\bibfnamefont#1{#1}\fi
\expandafter\ifx\csname citenamefont\endcsname\relax
  \def\citenamefont#1{#1}\fi
\expandafter\ifx\csname url\endcsname\relax
  \def\url#1{\texttt{#1}}\fi
\expandafter\ifx\csname urlprefix\endcsname\relax\def\urlprefix{URL }\fi
\providecommand{\bibinfo}[2]{#2}
\providecommand{\eprint}[2][]{\url{#2}}

\bibitem[{\citenamefont{Wendin}(2017)}]{Wendin_2017}
\bibinfo{author}{\bibfnamefont{G.}~\bibnamefont{Wendin}},
  \bibinfo{journal}{Reports on Progress in Physics}
  \textbf{\bibinfo{volume}{80}}, \bibinfo{pages}{106001}
  (\bibinfo{year}{2017}), ISSN \bibinfo{issn}{1361-6633}.

\bibitem[{\citenamefont{Slussarenko and Pryde}(2019)}]{Slussarenko_2019}
\bibinfo{author}{\bibfnamefont{S.}~\bibnamefont{Slussarenko}} \bibnamefont{and}
  \bibinfo{author}{\bibfnamefont{G.~J.} \bibnamefont{Pryde}},
  \bibinfo{journal}{Applied Physics Reviews} \textbf{\bibinfo{volume}{6}},
  \bibinfo{pages}{041303} (\bibinfo{year}{2019}), ISSN
  \bibinfo{issn}{1931-9401}.

\bibitem[{\citenamefont{Meher and Sivakumar}(2022)}]{Meher_2022}
\bibinfo{author}{\bibfnamefont{N.}~\bibnamefont{Meher}} \bibnamefont{and}
  \bibinfo{author}{\bibfnamefont{S.}~\bibnamefont{Sivakumar}},
  \bibinfo{journal}{The European Physical Journal Plus}
  \textbf{\bibinfo{volume}{137}}, \bibinfo{pages}{985} (\bibinfo{year}{2022}),
  ISSN \bibinfo{issn}{2190-5444}.

\bibitem[{\citenamefont{Degen et~al.}(2017)\citenamefont{Degen, Reinhard, and
  Cappellaro}}]{Degen_2017}
\bibinfo{author}{\bibfnamefont{C.}~\bibnamefont{Degen}},
  \bibinfo{author}{\bibfnamefont{F.}~\bibnamefont{Reinhard}}, \bibnamefont{and}
  \bibinfo{author}{\bibfnamefont{P.}~\bibnamefont{Cappellaro}},
  \bibinfo{journal}{Reviews of Modern Physics} \textbf{\bibinfo{volume}{89}}
  (\bibinfo{year}{2017}), ISSN \bibinfo{issn}{1539-0756}.

\bibitem[{\citenamefont{Polino et~al.}(2020)\citenamefont{Polino, Valeri,
  Spagnolo, and Sciarrino}}]{Polino_2020}
\bibinfo{author}{\bibfnamefont{E.}~\bibnamefont{Polino}},
  \bibinfo{author}{\bibfnamefont{M.}~\bibnamefont{Valeri}},
  \bibinfo{author}{\bibfnamefont{N.}~\bibnamefont{Spagnolo}}, \bibnamefont{and}
  \bibinfo{author}{\bibfnamefont{F.}~\bibnamefont{Sciarrino}},
  \bibinfo{journal}{{AVS} Quantum Science} \textbf{\bibinfo{volume}{2}},
  \bibinfo{pages}{024703} (\bibinfo{year}{2020}).

\bibitem[{\citenamefont{Stray et~al.}(2022)\citenamefont{Stray, Lamb, Kaushik,
  Vovrosh, Rodgers, Winch, Hayati, Boddice, Stabrawa, Niggebaum
  et~al.}}]{Stray_2022}
\bibinfo{author}{\bibfnamefont{B.}~\bibnamefont{Stray}},
  \bibinfo{author}{\bibfnamefont{A.}~\bibnamefont{Lamb}},
  \bibinfo{author}{\bibfnamefont{A.}~\bibnamefont{Kaushik}},
  \bibinfo{author}{\bibfnamefont{J.}~\bibnamefont{Vovrosh}},
  \bibinfo{author}{\bibfnamefont{A.}~\bibnamefont{Rodgers}},
  \bibinfo{author}{\bibfnamefont{J.}~\bibnamefont{Winch}},
  \bibinfo{author}{\bibfnamefont{F.}~\bibnamefont{Hayati}},
  \bibinfo{author}{\bibfnamefont{D.}~\bibnamefont{Boddice}},
  \bibinfo{author}{\bibfnamefont{A.}~\bibnamefont{Stabrawa}},
  \bibinfo{author}{\bibfnamefont{A.}~\bibnamefont{Niggebaum}},
  \bibnamefont{et~al.}, \bibinfo{journal}{Nature}
  \textbf{\bibinfo{volume}{602}}, \bibinfo{pages}{590} (\bibinfo{year}{2022}),
  ISSN \bibinfo{issn}{1476-4687}.

\bibitem[{\citenamefont{Barbieri}(2022)}]{Barbieri_2022}
\bibinfo{author}{\bibfnamefont{M.}~\bibnamefont{Barbieri}},
  \bibinfo{journal}{PRX Quantum} \textbf{\bibinfo{volume}{3}},
  \bibinfo{pages}{010202} (\bibinfo{year}{2022}).

\bibitem[{\citenamefont{Zidan et~al.}(2021)\citenamefont{Zidan, Eleuch, and
  Abdel-Aty}}]{Zidan_2021}
\bibinfo{author}{\bibfnamefont{M.}~\bibnamefont{Zidan}},
  \bibinfo{author}{\bibfnamefont{H.}~\bibnamefont{Eleuch}}, \bibnamefont{and}
  \bibinfo{author}{\bibfnamefont{M.}~\bibnamefont{Abdel-Aty}},
  \bibinfo{journal}{Results in Physics} \textbf{\bibinfo{volume}{21}},
  \bibinfo{pages}{103536} (\bibinfo{year}{2021}), ISSN
  \bibinfo{issn}{2211-3797}.

\bibitem[{\citenamefont{Arute et~al.}(2019)\citenamefont{Arute, Arya, Babbush,
  Bacon, Bardin, Barends, Biswas, Boixo, Brandao, Buell et~al.}}]{Arute_2019}
\bibinfo{author}{\bibfnamefont{F.}~\bibnamefont{Arute}},
  \bibinfo{author}{\bibfnamefont{K.}~\bibnamefont{Arya}},
  \bibinfo{author}{\bibfnamefont{R.}~\bibnamefont{Babbush}},
  \bibinfo{author}{\bibfnamefont{D.}~\bibnamefont{Bacon}},
  \bibinfo{author}{\bibfnamefont{J.~C.} \bibnamefont{Bardin}},
  \bibinfo{author}{\bibfnamefont{R.}~\bibnamefont{Barends}},
  \bibinfo{author}{\bibfnamefont{R.}~\bibnamefont{Biswas}},
  \bibinfo{author}{\bibfnamefont{S.}~\bibnamefont{Boixo}},
  \bibinfo{author}{\bibfnamefont{F.~G. S.~L.} \bibnamefont{Brandao}},
  \bibinfo{author}{\bibfnamefont{D.~A.} \bibnamefont{Buell}},
  \bibnamefont{et~al.}, \bibinfo{journal}{Nature}
  \textbf{\bibinfo{volume}{574}}, \bibinfo{pages}{505} (\bibinfo{year}{2019}),
  ISSN \bibinfo{issn}{1476-4687}.

\bibitem[{\citenamefont{Zhong et~al.}(2020)\citenamefont{Zhong, Wang, Deng,
  Chen, Peng, Luo, Qin, Wu, Ding, Hu et~al.}}]{Zhong_2020}
\bibinfo{author}{\bibfnamefont{H.-S.} \bibnamefont{Zhong}},
  \bibinfo{author}{\bibfnamefont{H.}~\bibnamefont{Wang}},
  \bibinfo{author}{\bibfnamefont{Y.-H.} \bibnamefont{Deng}},
  \bibinfo{author}{\bibfnamefont{M.-C.} \bibnamefont{Chen}},
  \bibinfo{author}{\bibfnamefont{L.-C.} \bibnamefont{Peng}},
  \bibinfo{author}{\bibfnamefont{Y.-H.} \bibnamefont{Luo}},
  \bibinfo{author}{\bibfnamefont{J.}~\bibnamefont{Qin}},
  \bibinfo{author}{\bibfnamefont{D.}~\bibnamefont{Wu}},
  \bibinfo{author}{\bibfnamefont{X.}~\bibnamefont{Ding}},
  \bibinfo{author}{\bibfnamefont{Y.}~\bibnamefont{Hu}}, \bibnamefont{et~al.},
  \bibinfo{journal}{Science} \textbf{\bibinfo{volume}{370}},
  \bibinfo{pages}{1460} (\bibinfo{year}{2020}), ISSN \bibinfo{issn}{1095-9203}.

\bibitem[{\citenamefont{Zhong et~al.}(2021)\citenamefont{Zhong, Deng, Qin,
  Wang, Chen, Peng, Luo, Wu, Gong, Su et~al.}}]{Zhong_2021}
\bibinfo{author}{\bibfnamefont{H.-S.} \bibnamefont{Zhong}},
  \bibinfo{author}{\bibfnamefont{Y.-H.} \bibnamefont{Deng}},
  \bibinfo{author}{\bibfnamefont{J.}~\bibnamefont{Qin}},
  \bibinfo{author}{\bibfnamefont{H.}~\bibnamefont{Wang}},
  \bibinfo{author}{\bibfnamefont{M.-C.} \bibnamefont{Chen}},
  \bibinfo{author}{\bibfnamefont{L.-C.} \bibnamefont{Peng}},
  \bibinfo{author}{\bibfnamefont{Y.-H.} \bibnamefont{Luo}},
  \bibinfo{author}{\bibfnamefont{D.}~\bibnamefont{Wu}},
  \bibinfo{author}{\bibfnamefont{S.-Q.} \bibnamefont{Gong}},
  \bibinfo{author}{\bibfnamefont{H.}~\bibnamefont{Su}}, \bibnamefont{et~al.},
  \bibinfo{journal}{Physical Review Letters} \textbf{\bibinfo{volume}{127}}
  (\bibinfo{year}{2021}), ISSN \bibinfo{issn}{1079-7114}.

\bibitem[{\citenamefont{Aspelmeyer et~al.}(2014)\citenamefont{Aspelmeyer,
  Kippenberg, and Marquardt}}]{Aspelmeyer.2014}
\bibinfo{author}{\bibfnamefont{M.}~\bibnamefont{Aspelmeyer}},
  \bibinfo{author}{\bibfnamefont{T.~J.} \bibnamefont{Kippenberg}},
  \bibnamefont{and}
  \bibinfo{author}{\bibfnamefont{F.}~\bibnamefont{Marquardt}},
  \bibinfo{journal}{Reviews of Modern Physics} \textbf{\bibinfo{volume}{86}},
  \bibinfo{pages}{1391} (\bibinfo{year}{2014}), ISSN \bibinfo{issn}{1539-0756}.

\bibitem[{\citenamefont{Foulla et~al.}(2017)\citenamefont{Foulla, Djorw{\'{e}},
  Kingni, and Engo}}]{Foulla_2017}
\bibinfo{author}{\bibfnamefont{D.~P.} \bibnamefont{Foulla}},
  \bibinfo{author}{\bibfnamefont{P.}~\bibnamefont{Djorw{\'{e}}}},
  \bibinfo{author}{\bibfnamefont{S.~T.} \bibnamefont{Kingni}},
  \bibnamefont{and} \bibinfo{author}{\bibfnamefont{S.~G.~N.}
  \bibnamefont{Engo}}, \bibinfo{journal}{Physical Review A}
  \textbf{\bibinfo{volume}{95}}, \bibinfo{pages}{123} (\bibinfo{year}{2017}).

\bibitem[{\citenamefont{Riedinger et~al.}(2018)\citenamefont{Riedinger,
  Wallucks, Marinković, Löschnauer, Aspelmeyer, Hong, and
  Gröblacher}}]{Riedinger_2018}
\bibinfo{author}{\bibfnamefont{R.}~\bibnamefont{Riedinger}},
  \bibinfo{author}{\bibfnamefont{A.}~\bibnamefont{Wallucks}},
  \bibinfo{author}{\bibfnamefont{I.}~\bibnamefont{Marinković}},
  \bibinfo{author}{\bibfnamefont{C.}~\bibnamefont{Löschnauer}},
  \bibinfo{author}{\bibfnamefont{M.}~\bibnamefont{Aspelmeyer}},
  \bibinfo{author}{\bibfnamefont{S.}~\bibnamefont{Hong}}, \bibnamefont{and}
  \bibinfo{author}{\bibfnamefont{S.}~\bibnamefont{Gröblacher}},
  \bibinfo{journal}{Nature} \textbf{\bibinfo{volume}{556}},
  \bibinfo{pages}{473} (\bibinfo{year}{2018}), ISSN \bibinfo{issn}{1476-4687}.

\bibitem[{\citenamefont{Tchodimou et~al.}(2017)\citenamefont{Tchodimou, Djorwe,
  and Engo}}]{Tchodimou.2017}
\bibinfo{author}{\bibfnamefont{C.}~\bibnamefont{Tchodimou}},
  \bibinfo{author}{\bibfnamefont{P.}~\bibnamefont{Djorwe}}, \bibnamefont{and}
  \bibinfo{author}{\bibfnamefont{S.~G.~N.} \bibnamefont{Engo}},
  \bibinfo{journal}{Physical Review A} \textbf{\bibinfo{volume}{96}}
  (\bibinfo{year}{2017}).

\bibitem[{\citenamefont{Kotler et~al.}(2021)\citenamefont{Kotler, Peterson,
  Shojaee, Lecocq, Cicak, Kwiatkowski, Geller, Glancy, Knill, Simmonds
  et~al.}}]{Kotler_2021}
\bibinfo{author}{\bibfnamefont{S.}~\bibnamefont{Kotler}},
  \bibinfo{author}{\bibfnamefont{G.~A.} \bibnamefont{Peterson}},
  \bibinfo{author}{\bibfnamefont{E.}~\bibnamefont{Shojaee}},
  \bibinfo{author}{\bibfnamefont{F.}~\bibnamefont{Lecocq}},
  \bibinfo{author}{\bibfnamefont{K.}~\bibnamefont{Cicak}},
  \bibinfo{author}{\bibfnamefont{A.}~\bibnamefont{Kwiatkowski}},
  \bibinfo{author}{\bibfnamefont{S.}~\bibnamefont{Geller}},
  \bibinfo{author}{\bibfnamefont{S.}~\bibnamefont{Glancy}},
  \bibinfo{author}{\bibfnamefont{E.}~\bibnamefont{Knill}},
  \bibinfo{author}{\bibfnamefont{R.~W.} \bibnamefont{Simmonds}},
  \bibnamefont{et~al.}, \bibinfo{journal}{Science}
  \textbf{\bibinfo{volume}{372}}, \bibinfo{pages}{622} (\bibinfo{year}{2021}),
  ISSN \bibinfo{issn}{1095-9203}.

\bibitem[{\citenamefont{Chakraborty and Sarma}(2019)}]{Chakrabo.2019}
\bibinfo{author}{\bibfnamefont{S.}~\bibnamefont{Chakraborty}} \bibnamefont{and}
  \bibinfo{author}{\bibfnamefont{A.~K.} \bibnamefont{Sarma}},
  \bibinfo{journal}{Physical Review A} \textbf{\bibinfo{volume}{100}},
  \bibinfo{pages}{063846} (\bibinfo{year}{2019}), ISSN
  \bibinfo{issn}{2469-9934}.

\bibitem[{\citenamefont{Massembele et~al.}(2024)\citenamefont{Massembele,
  Djorwé, Sarma, and Engo}}]{Rostand1.2024}
\bibinfo{author}{\bibfnamefont{D.~R.~K.} \bibnamefont{Massembele}},
  \bibinfo{author}{\bibfnamefont{P.}~\bibnamefont{Djorwé}},
  \bibinfo{author}{\bibfnamefont{A.~K.} \bibnamefont{Sarma}}, \bibnamefont{and}
  \bibinfo{author}{\bibfnamefont{S.~G.~N.} \bibnamefont{Engo}},
  \bibinfo{journal}{arXiv}  (\bibinfo{year}{2024}).

\bibitem[{\citenamefont{Xia et~al.}(2023)\citenamefont{Xia, Agrawal, Pluchar,
  Brady, Liu, Zhuang, Wilson, and Zhang}}]{Xia_2023}
\bibinfo{author}{\bibfnamefont{Y.}~\bibnamefont{Xia}},
  \bibinfo{author}{\bibfnamefont{A.~R.} \bibnamefont{Agrawal}},
  \bibinfo{author}{\bibfnamefont{C.~M.} \bibnamefont{Pluchar}},
  \bibinfo{author}{\bibfnamefont{A.~J.} \bibnamefont{Brady}},
  \bibinfo{author}{\bibfnamefont{Z.}~\bibnamefont{Liu}},
  \bibinfo{author}{\bibfnamefont{Q.}~\bibnamefont{Zhuang}},
  \bibinfo{author}{\bibfnamefont{D.~J.} \bibnamefont{Wilson}},
  \bibnamefont{and} \bibinfo{author}{\bibfnamefont{Z.}~\bibnamefont{Zhang}},
  \bibinfo{journal}{Nature Photonics} \textbf{\bibinfo{volume}{17}},
  \bibinfo{pages}{470} (\bibinfo{year}{2023}), ISSN \bibinfo{issn}{1749-4893}.

\bibitem[{\citenamefont{Brady et~al.}(2023)\citenamefont{Brady, Chen, Xia,
  Manley, Dey~Chowdhury, Xiao, Liu, Harnik, Wilson, Zhang et~al.}}]{Brady_2023}
\bibinfo{author}{\bibfnamefont{A.~J.} \bibnamefont{Brady}},
  \bibinfo{author}{\bibfnamefont{X.}~\bibnamefont{Chen}},
  \bibinfo{author}{\bibfnamefont{Y.}~\bibnamefont{Xia}},
  \bibinfo{author}{\bibfnamefont{J.}~\bibnamefont{Manley}},
  \bibinfo{author}{\bibfnamefont{M.}~\bibnamefont{Dey~Chowdhury}},
  \bibinfo{author}{\bibfnamefont{K.}~\bibnamefont{Xiao}},
  \bibinfo{author}{\bibfnamefont{Z.}~\bibnamefont{Liu}},
  \bibinfo{author}{\bibfnamefont{R.}~\bibnamefont{Harnik}},
  \bibinfo{author}{\bibfnamefont{D.~J.} \bibnamefont{Wilson}},
  \bibinfo{author}{\bibfnamefont{Z.}~\bibnamefont{Zhang}},
  \bibnamefont{et~al.}, \bibinfo{journal}{Communications Physics}
  \textbf{\bibinfo{volume}{6}} (\bibinfo{year}{2023}), ISSN
  \bibinfo{issn}{2399-3650}.

\bibitem[{\citenamefont{Vahlbruch et~al.}(2016)\citenamefont{Vahlbruch, Mehmet,
  Danzmann, and Schnabel}}]{Vahlbruch_2016}
\bibinfo{author}{\bibfnamefont{H.}~\bibnamefont{Vahlbruch}},
  \bibinfo{author}{\bibfnamefont{M.}~\bibnamefont{Mehmet}},
  \bibinfo{author}{\bibfnamefont{K.}~\bibnamefont{Danzmann}}, \bibnamefont{and}
  \bibinfo{author}{\bibfnamefont{R.}~\bibnamefont{Schnabel}},
  \bibinfo{journal}{Physical Review Letters} \textbf{\bibinfo{volume}{117}}
  (\bibinfo{year}{2016}), ISSN \bibinfo{issn}{1079-7114}.

\bibitem[{\citenamefont{Qin et~al.}(2022)\citenamefont{Qin, Miranowicz, and
  Nori}}]{Qin_2022}
\bibinfo{author}{\bibfnamefont{W.}~\bibnamefont{Qin}},
  \bibinfo{author}{\bibfnamefont{A.}~\bibnamefont{Miranowicz}},
  \bibnamefont{and} \bibinfo{author}{\bibfnamefont{F.}~\bibnamefont{Nori}},
  \bibinfo{journal}{Physical Review Letters} \textbf{\bibinfo{volume}{129}}
  (\bibinfo{year}{2022}), ISSN \bibinfo{issn}{1079-7114}.

\bibitem[{\citenamefont{Clark et~al.}(2017)\citenamefont{Clark, Lecocq,
  Simmonds, Aumentado, and Teufel}}]{Clark_2017}
\bibinfo{author}{\bibfnamefont{J.~B.} \bibnamefont{Clark}},
  \bibinfo{author}{\bibfnamefont{F.}~\bibnamefont{Lecocq}},
  \bibinfo{author}{\bibfnamefont{R.~W.} \bibnamefont{Simmonds}},
  \bibinfo{author}{\bibfnamefont{J.}~\bibnamefont{Aumentado}},
  \bibnamefont{and} \bibinfo{author}{\bibfnamefont{J.~D.}
  \bibnamefont{Teufel}}, \bibinfo{journal}{Nature}
  \textbf{\bibinfo{volume}{541}}, \bibinfo{pages}{191} (\bibinfo{year}{2017}),
  ISSN \bibinfo{issn}{1476-4687}.

\bibitem[{\citenamefont{Djorw{\'{e}} et~al.}(2012)\citenamefont{Djorw{\'{e}},
  Mb{\'{e}}, Engo, and Woafo}}]{Djor.2012}
\bibinfo{author}{\bibfnamefont{P.}~\bibnamefont{Djorw{\'{e}}}},
  \bibinfo{author}{\bibfnamefont{J.~H.~T.} \bibnamefont{Mb{\'{e}}}},
  \bibinfo{author}{\bibfnamefont{S.~G.~N.} \bibnamefont{Engo}},
  \bibnamefont{and} \bibinfo{author}{\bibfnamefont{P.}~\bibnamefont{Woafo}},
  \bibinfo{journal}{Physical Review A} \textbf{\bibinfo{volume}{86}},
  \bibinfo{pages}{043816} (\bibinfo{year}{2012}).

\bibitem[{\citenamefont{Aasi et~al.}(2013)\citenamefont{Aasi, Abadie, Abbott,
  Abbott, Abbott, Abernathy, Adams, Adams, Addesso, Adhikari
  et~al.}}]{Aasi_2013}
\bibinfo{author}{\bibfnamefont{J.}~\bibnamefont{Aasi}},
  \bibinfo{author}{\bibfnamefont{J.}~\bibnamefont{Abadie}},
  \bibinfo{author}{\bibfnamefont{B.~P.} \bibnamefont{Abbott}},
  \bibinfo{author}{\bibfnamefont{R.}~\bibnamefont{Abbott}},
  \bibinfo{author}{\bibfnamefont{T.~D.} \bibnamefont{Abbott}},
  \bibinfo{author}{\bibfnamefont{M.~R.} \bibnamefont{Abernathy}},
  \bibinfo{author}{\bibfnamefont{C.}~\bibnamefont{Adams}},
  \bibinfo{author}{\bibfnamefont{T.}~\bibnamefont{Adams}},
  \bibinfo{author}{\bibfnamefont{P.}~\bibnamefont{Addesso}},
  \bibinfo{author}{\bibfnamefont{R.~X.} \bibnamefont{Adhikari}},
  \bibnamefont{et~al.}, \bibinfo{journal}{Nature Photonics}
  \textbf{\bibinfo{volume}{7}}, \bibinfo{pages}{613} (\bibinfo{year}{2013}),
  ISSN \bibinfo{issn}{1749-4893}.

\bibitem[{\citenamefont{et~al.}(2016)}]{Abbott.2016}
\bibinfo{author}{\bibfnamefont{B.~P.~A.} \bibnamefont{et~al.}},
  \bibinfo{journal}{Physical Review Letters} \textbf{\bibinfo{volume}{116}},
  \bibinfo{pages}{061102} (\bibinfo{year}{2016}).

\bibitem[{\citenamefont{Qin et~al.}(2021)\citenamefont{Qin, Miranowicz, Jing,
  and Nori}}]{Qin_2021}
\bibinfo{author}{\bibfnamefont{W.}~\bibnamefont{Qin}},
  \bibinfo{author}{\bibfnamefont{A.}~\bibnamefont{Miranowicz}},
  \bibinfo{author}{\bibfnamefont{H.}~\bibnamefont{Jing}}, \bibnamefont{and}
  \bibinfo{author}{\bibfnamefont{F.}~\bibnamefont{Nori}},
  \bibinfo{journal}{Physical Review Letters} \textbf{\bibinfo{volume}{127}}
  (\bibinfo{year}{2021}), ISSN \bibinfo{issn}{1079-7114}.

\bibitem[{\citenamefont{Banerjee et~al.}(2023)\citenamefont{Banerjee, Kalita,
  and Sarma}}]{Banerjee_2023}
\bibinfo{author}{\bibfnamefont{P.}~\bibnamefont{Banerjee}},
  \bibinfo{author}{\bibfnamefont{S.}~\bibnamefont{Kalita}}, \bibnamefont{and}
  \bibinfo{author}{\bibfnamefont{A.~K.} \bibnamefont{Sarma}},
  \bibinfo{journal}{Journal of the Optical Society of America B}
  \textbf{\bibinfo{volume}{40}}, \bibinfo{pages}{1398} (\bibinfo{year}{2023}),
  ISSN \bibinfo{issn}{1520-8540}.

\bibitem[{\citenamefont{Djorwe et~al.}(2019)\citenamefont{Djorwe, Pennec, and
  Djafari-Rouhani}}]{Djorwe.2019}
\bibinfo{author}{\bibfnamefont{P.}~\bibnamefont{Djorwe}},
  \bibinfo{author}{\bibfnamefont{Y.}~\bibnamefont{Pennec}}, \bibnamefont{and}
  \bibinfo{author}{\bibfnamefont{B.}~\bibnamefont{Djafari-Rouhani}},
  \bibinfo{journal}{Physical Review Applied} \textbf{\bibinfo{volume}{12}},
  \bibinfo{pages}{024002} (\bibinfo{year}{2019}).

\bibitem[{\citenamefont{Tchounda et~al.}(2023)\citenamefont{Tchounda, Djorwé,
  Engo, and Djafari-Rouhani}}]{Tchounda_2023}
\bibinfo{author}{\bibfnamefont{S.~M.} \bibnamefont{Tchounda}},
  \bibinfo{author}{\bibfnamefont{P.}~\bibnamefont{Djorwé}},
  \bibinfo{author}{\bibfnamefont{S.~N.} \bibnamefont{Engo}}, \bibnamefont{and}
  \bibinfo{author}{\bibfnamefont{B.}~\bibnamefont{Djafari-Rouhani}},
  \bibinfo{journal}{Physical Review Applied} \textbf{\bibinfo{volume}{19}},
  \bibinfo{pages}{064016} (\bibinfo{year}{2023}), ISSN
  \bibinfo{issn}{2331-7019}.

\bibitem[{\citenamefont{Djorwé
  et~al.}(2023{\natexlab{a}})\citenamefont{Djorwé, Asjad, Pennec, Dutykh, and
  Djafari-Rouhani}}]{djorwe.PRR}
\bibinfo{author}{\bibfnamefont{P.}~\bibnamefont{Djorwé}},
  \bibinfo{author}{\bibfnamefont{M.}~\bibnamefont{Asjad}},
  \bibinfo{author}{\bibfnamefont{Y.}~\bibnamefont{Pennec}},
  \bibinfo{author}{\bibfnamefont{D.}~\bibnamefont{Dutykh}}, \bibnamefont{and}
  \bibinfo{author}{\bibfnamefont{B.}~\bibnamefont{Djafari-Rouhani}},
  \bibinfo{journal}{arXiv}  (\bibinfo{year}{2023}{\natexlab{a}}).

\bibitem[{\citenamefont{Stannigel et~al.}(2012)\citenamefont{Stannigel, Komar,
  Habraken, Bennett, Lukin, Zoller, and Rabl}}]{Stannigel_2012}
\bibinfo{author}{\bibfnamefont{K.}~\bibnamefont{Stannigel}},
  \bibinfo{author}{\bibfnamefont{P.}~\bibnamefont{Komar}},
  \bibinfo{author}{\bibfnamefont{S.~J.~M.} \bibnamefont{Habraken}},
  \bibinfo{author}{\bibfnamefont{S.~D.} \bibnamefont{Bennett}},
  \bibinfo{author}{\bibfnamefont{M.~D.} \bibnamefont{Lukin}},
  \bibinfo{author}{\bibfnamefont{P.}~\bibnamefont{Zoller}}, \bibnamefont{and}
  \bibinfo{author}{\bibfnamefont{P.}~\bibnamefont{Rabl}},
  \bibinfo{journal}{Physical Review Letters} \textbf{\bibinfo{volume}{109}}
  (\bibinfo{year}{2012}), ISSN \bibinfo{issn}{1079-7114}.

\bibitem[{\citenamefont{Madiot et~al.}(2023)\citenamefont{Madiot, Ng, Arregui,
  Florez, Albrechtsen, Stobbe, García, and Sotomayor-Torres}}]{Madiot_2023}
\bibinfo{author}{\bibfnamefont{G.}~\bibnamefont{Madiot}},
  \bibinfo{author}{\bibfnamefont{R.~C.} \bibnamefont{Ng}},
  \bibinfo{author}{\bibfnamefont{G.}~\bibnamefont{Arregui}},
  \bibinfo{author}{\bibfnamefont{O.}~\bibnamefont{Florez}},
  \bibinfo{author}{\bibfnamefont{M.}~\bibnamefont{Albrechtsen}},
  \bibinfo{author}{\bibfnamefont{S.}~\bibnamefont{Stobbe}},
  \bibinfo{author}{\bibfnamefont{P.~D.} \bibnamefont{García}},
  \bibnamefont{and} \bibinfo{author}{\bibfnamefont{C.~M.}
  \bibnamefont{Sotomayor-Torres}}, \bibinfo{journal}{Physical Review Letters}
  \textbf{\bibinfo{volume}{130}} (\bibinfo{year}{2023}), ISSN
  \bibinfo{issn}{1079-7114}.

\bibitem[{\citenamefont{Ren et~al.}(2022)\citenamefont{Ren, Shah, Pfeifer,
  Brendel, Peano, Marquardt, and Painter}}]{Ren_2022}
\bibinfo{author}{\bibfnamefont{H.}~\bibnamefont{Ren}},
  \bibinfo{author}{\bibfnamefont{T.}~\bibnamefont{Shah}},
  \bibinfo{author}{\bibfnamefont{H.}~\bibnamefont{Pfeifer}},
  \bibinfo{author}{\bibfnamefont{C.}~\bibnamefont{Brendel}},
  \bibinfo{author}{\bibfnamefont{V.}~\bibnamefont{Peano}},
  \bibinfo{author}{\bibfnamefont{F.}~\bibnamefont{Marquardt}},
  \bibnamefont{and} \bibinfo{author}{\bibfnamefont{O.}~\bibnamefont{Painter}},
  \bibinfo{journal}{Nature Communications} \textbf{\bibinfo{volume}{13}}
  (\bibinfo{year}{2022}), ISSN \bibinfo{issn}{2041-1723}.

\bibitem[{\citenamefont{Woolley and Clerk}(2014)}]{Woolley_2014}
\bibinfo{author}{\bibfnamefont{M.~J.} \bibnamefont{Woolley}} \bibnamefont{and}
  \bibinfo{author}{\bibfnamefont{A.~A.} \bibnamefont{Clerk}},
  \bibinfo{journal}{Physical Review A} \textbf{\bibinfo{volume}{89}}
  (\bibinfo{year}{2014}), ISSN \bibinfo{issn}{1094-1622}.

\bibitem[{\citenamefont{Wollman et~al.}(2015)\citenamefont{Wollman, Lei,
  Weinstein, Suh, Kronwald, Marquardt, Clerk, and Schwab}}]{Wollman_2015}
\bibinfo{author}{\bibfnamefont{E.~E.} \bibnamefont{Wollman}},
  \bibinfo{author}{\bibfnamefont{C.~U.} \bibnamefont{Lei}},
  \bibinfo{author}{\bibfnamefont{A.~J.} \bibnamefont{Weinstein}},
  \bibinfo{author}{\bibfnamefont{J.}~\bibnamefont{Suh}},
  \bibinfo{author}{\bibfnamefont{A.}~\bibnamefont{Kronwald}},
  \bibinfo{author}{\bibfnamefont{F.}~\bibnamefont{Marquardt}},
  \bibinfo{author}{\bibfnamefont{A.~A.} \bibnamefont{Clerk}}, \bibnamefont{and}
  \bibinfo{author}{\bibfnamefont{K.~C.} \bibnamefont{Schwab}},
  \bibinfo{journal}{Science} \textbf{\bibinfo{volume}{349}},
  \bibinfo{pages}{952} (\bibinfo{year}{2015}), ISSN \bibinfo{issn}{1095-9203}.

\bibitem[{\citenamefont{Lei et~al.}(2016)\citenamefont{Lei, Weinstein, Suh,
  Wollman, Kronwald, Marquardt, Clerk, and Schwab}}]{Lei_2016}
\bibinfo{author}{\bibfnamefont{C.}~\bibnamefont{Lei}},
  \bibinfo{author}{\bibfnamefont{A.}~\bibnamefont{Weinstein}},
  \bibinfo{author}{\bibfnamefont{J.}~\bibnamefont{Suh}},
  \bibinfo{author}{\bibfnamefont{E.}~\bibnamefont{Wollman}},
  \bibinfo{author}{\bibfnamefont{A.}~\bibnamefont{Kronwald}},
  \bibinfo{author}{\bibfnamefont{F.}~\bibnamefont{Marquardt}},
  \bibinfo{author}{\bibfnamefont{A.}~\bibnamefont{Clerk}}, \bibnamefont{and}
  \bibinfo{author}{\bibfnamefont{K.}~\bibnamefont{Schwab}},
  \bibinfo{journal}{Physical Review Letters} \textbf{\bibinfo{volume}{117}}
  (\bibinfo{year}{2016}), ISSN \bibinfo{issn}{1079-7114}.

\bibitem[{\citenamefont{Schmidt et~al.}(2015)\citenamefont{Schmidt, Kessler,
  Peano, Painter, and Marquardt}}]{Schmidt.2015}
\bibinfo{author}{\bibfnamefont{M.}~\bibnamefont{Schmidt}},
  \bibinfo{author}{\bibfnamefont{S.}~\bibnamefont{Kessler}},
  \bibinfo{author}{\bibfnamefont{V.}~\bibnamefont{Peano}},
  \bibinfo{author}{\bibfnamefont{O.}~\bibnamefont{Painter}}, \bibnamefont{and}
  \bibinfo{author}{\bibfnamefont{F.}~\bibnamefont{Marquardt}},
  \bibinfo{journal}{Optica} \textbf{\bibinfo{volume}{2}}, \bibinfo{pages}{635}
  (\bibinfo{year}{2015}).

\bibitem[{\citenamefont{Fang et~al.}(2012)\citenamefont{Fang, Yu, and
  Fan}}]{Fang_2012}
\bibinfo{author}{\bibfnamefont{K.}~\bibnamefont{Fang}},
  \bibinfo{author}{\bibfnamefont{Z.}~\bibnamefont{Yu}}, \bibnamefont{and}
  \bibinfo{author}{\bibfnamefont{S.}~\bibnamefont{Fan}},
  \bibinfo{journal}{Nature Photonics} \textbf{\bibinfo{volume}{6}},
  \bibinfo{pages}{782} (\bibinfo{year}{2012}).

\bibitem[{\citenamefont{Brendel et~al.}(2017)\citenamefont{Brendel, Peano,
  Painter, and Marquardt}}]{Brendel.2017}
\bibinfo{author}{\bibfnamefont{C.}~\bibnamefont{Brendel}},
  \bibinfo{author}{\bibfnamefont{V.}~\bibnamefont{Peano}},
  \bibinfo{author}{\bibfnamefont{O.~J.} \bibnamefont{Painter}},
  \bibnamefont{and}
  \bibinfo{author}{\bibfnamefont{F.}~\bibnamefont{Marquardt}},
  \bibinfo{journal}{Proceedings of the National Academy of Sciences}
  \textbf{\bibinfo{volume}{114}} (\bibinfo{year}{2017}).

\bibitem[{\citenamefont{Mathew et~al.}(2020)\citenamefont{Mathew, del Pino, and
  Verhagen}}]{Mathew_2020}
\bibinfo{author}{\bibfnamefont{J.~P.} \bibnamefont{Mathew}},
  \bibinfo{author}{\bibfnamefont{J.}~\bibnamefont{del Pino}}, \bibnamefont{and}
  \bibinfo{author}{\bibfnamefont{E.}~\bibnamefont{Verhagen}},
  \bibinfo{journal}{Nature Nanotechnology} \textbf{\bibinfo{volume}{15}},
  \bibinfo{pages}{198} (\bibinfo{year}{2020}).

\bibitem[{\citenamefont{Wang et~al.}(2020)\citenamefont{Wang, Li, and
  Li}}]{Wang_2020}
\bibinfo{author}{\bibfnamefont{X.}~\bibnamefont{Wang}},
  \bibinfo{author}{\bibfnamefont{H.-R.} \bibnamefont{Li}}, \bibnamefont{and}
  \bibinfo{author}{\bibfnamefont{F.-L.} \bibnamefont{Li}},
  \bibinfo{journal}{New Journal of Physics} \textbf{\bibinfo{volume}{22}},
  \bibinfo{pages}{033037} (\bibinfo{year}{2020}).

\bibitem[{\citenamefont{Lai et~al.}(2022)\citenamefont{Lai, Liao, Miranowicz,
  and Nori}}]{Lai_2022}
\bibinfo{author}{\bibfnamefont{D.-G.} \bibnamefont{Lai}},
  \bibinfo{author}{\bibfnamefont{J.-Q.} \bibnamefont{Liao}},
  \bibinfo{author}{\bibfnamefont{A.}~\bibnamefont{Miranowicz}},
  \bibnamefont{and} \bibinfo{author}{\bibfnamefont{F.}~\bibnamefont{Nori}},
  \bibinfo{journal}{Physical Review Letters} \textbf{\bibinfo{volume}{129}},
  \bibinfo{pages}{063602} (\bibinfo{year}{2022}).

\bibitem[{\citenamefont{Shen et~al.}(2021)\citenamefont{Shen, Zhang, Zou, Guo,
  and Dong}}]{BSBS_basis}
\bibinfo{author}{\bibfnamefont{Z.}~\bibnamefont{Shen}},
  \bibinfo{author}{\bibfnamefont{Y.-L.} \bibnamefont{Zhang}},
  \bibinfo{author}{\bibfnamefont{C.-L.} \bibnamefont{Zou}},
  \bibinfo{author}{\bibfnamefont{G.-C.} \bibnamefont{Guo}}, \bibnamefont{and}
  \bibinfo{author}{\bibfnamefont{C.-H.} \bibnamefont{Dong}},
  \bibinfo{journal}{Phys. Rev. Lett.} \textbf{\bibinfo{volume}{126}},
  \bibinfo{pages}{163604} (\bibinfo{year}{2021}).

\bibitem[{\citenamefont{Djorwé
  et~al.}(2023{\natexlab{b}})\citenamefont{Djorwé, Alphonse, Abbagari, Doka,
  and Engo}}]{Djor.2023}
\bibinfo{author}{\bibfnamefont{P.}~\bibnamefont{Djorwé}},
  \bibinfo{author}{\bibfnamefont{H.}~\bibnamefont{Alphonse}},
  \bibinfo{author}{\bibfnamefont{S.}~\bibnamefont{Abbagari}},
  \bibinfo{author}{\bibfnamefont{S.}~\bibnamefont{Doka}}, \bibnamefont{and}
  \bibinfo{author}{\bibfnamefont{S.~N.} \bibnamefont{Engo}},
  \bibinfo{journal}{Chaos, Solitons and Fractals}
  \textbf{\bibinfo{volume}{170}}, \bibinfo{pages}{113333}
  (\bibinfo{year}{2023}{\natexlab{b}}), ISSN \bibinfo{issn}{0960-0779}.

\end{thebibliography}

\end{document}